\documentclass[11pt,twoside,letterpaper]{article} % Specifies the document style.
\usepackage{times,fancyhdr}
\usepackage[dvips]{color}
\usepackage{epsfig}
\usepackage{graphicx}
\usepackage{geometry}
\usepackage{titlesec}

\titlelabel{\thetitle.\quad}                                % Replace the colon that normally appears after Section numbers by a period.[Needs 'titlesec'  package!]

\usepackage[labelsep=period]{caption}                       % Replace the colon that normally appears after the Figure and Table number by a period.

\makeatletter 
\def\cleardoublepage{\clearpage\if@twoside \ifodd\c@page\else% 
    \hbox{}% 
    \thispagestyle{empty}% % Empty header styles 
    \newpage% 
    \if@twocolumn\hbox{}\newpage\fi\fi\fi} 
\makeatother 

\usepackage{amssymb}
\usepackage{amsmath}
\usepackage{amsfonts}
\begin{document}
\title{
{\begin{flushleft}
\vskip 0.45in
{\normalsize\bfseries\textit{Chapter~}}
\end{flushleft}
\vskip 0.45in
\bfseries\scshape Understanding Quaternions from Modern Algebra \\and Theoretical Physics}}
\author{\bfseries\itshape Sadataka Furui\thanks{E-mail address: furui@umb.teikyo-u.ac.jp.}\\
(formerly) Teikyo University,\\
 Graduate School of Science and Engineering,\\
Utsunomiya, Tochigi, Japan}
\date{}
\maketitle
\thispagestyle{empty}
\setcounter{page}{1}
%% ------------ [First Page Running Head] -------------------------------------------------
%\thispagestyle{fancy}
%\fancyhead{}
%\fancyhead[L]{In: xxxxxxxxxxxxxxx\\ 
%Editor: xxxxxxxxxxxxx, pp. {\thepage-\pageref{lastpage-01}}}
%\fancyhead[R]{ISBN: xxxxxxxxxxxxx\\
%\copyright~2020 Nova Science Publishers, Inc.}
%\fancyfoot{}
%\renewcommand{\headrulewidth}{0pt}
%%-------------------------------------------------------------------------------

%\begin{center}
%{\bf ABSTRACT}
%\end{center} 

\begin{abstract}
 Quaternions were appeared through Lagrangian formulation of mechanics in Symplectic vector space.  Its general form was obtained from the Clifford algebra, and Frobenius' theorem, which says that `` the only finite-dimensional real division algebra are the real field ${\bf R}$, the complex field ${\bf C}$ and the algebra ${\bf H}$ of quaternions'' was derived. 
They appear also through Hamilton formulation of mechanics, as elements of rotation groups in the symplectic vector spaces.

Quaternions were used in the solution of 4-dimensional Dirac equation in QED, and also in solutions of Yang-Mills equation in QCD as elements of noncommutative geometry. 

We present how quaternions are formulated in Clifford Algebra, how it is used in explaining rotation group in symplectic vector space and parallel transformation in holonomic dynamics. When a dynamical system has hysteresis, pre-symplectic manifolds and nonholonomic dynamics appear.

Quaternions represent rotation of 3-dimensional sphere ${\bf S}^3$. Artin's generalized quaternions and Rohlin-Pontryagin's embedding of quaternions on 4-dimensional manifolds, and Kodaira's embedding of quaternions on ${\bf S}^1\times {\bf S}^3$ manifolds are also discussed.
\end{abstract}
%\vspace{0.1in}


\vspace{.08in} 
\noindent \textbf{Keywords:} Quaternions, Frobenius' theorem, Clifford algebra, Symplectic Vector Space, Holonomy Groups
%% Other situations:
%\noindent \textbf{Key Words}: Gluon, Color and Flavor degree of freedom 
%\vspace{.08in} 
%\noindent {\textbf AMS Subject Classification:} %53D, 37C, 65P.
%% ------------ [Running Heads - for odd and even pages] - please insert it only on page 2!
%\pagestyle{fancy}
%\fancyhead{}
%%\fancyhead[EC]{Furui, Sadataka}
%%\fancyhead[EL,OR]{\thepage}
%%\fancyhead[OC]{Understanding Quaternions}
%\fancyfoot{}
%\renewcommand\headrulewidth{0.5pt} 
%%------------------------------------------------------------------------------

% ------------ [Running Heads - must be located on page 2!!!] -------------------------------------------------
\pagestyle{fancy}
\fancyhead{}
\fancyhead[EC]{\it Sadataka Furui}
\fancyhead[EL,OR]{\thepage}
\fancyhead[OC]{\it Understanding Quaternions from Modern Algebra ...}
\fancyfoot{}
\renewcommand\headrulewidth{0pt} 
%-------------------------------------------------------------------------------

\section{\protect\centering{\sc Introduction}}
Quaternion $\bf H$ was discovered by Hamilton \cite{Hamilton}, and known as one real division algebra like real numbers ${\bf R}$ and complex number ${\bf C}$ which satisfies a quadratic equation $x^2=\alpha e+\beta x$ ($\alpha,\beta\in{\bf R}$, $e$ is unit element) \cite{KR90a}.
Let $Q_n$ denotes $n$ dimensional vector space of $\bf{ R, C, H}$, and $G_n$ denotes the group of linear transformation that preserve inner products
\[
y_i=\sum_{i=1}^n \sigma_{ij} x_j.
\]
The condition $\sigma(x)\cdot \sigma(y)=x\cdot y$ leads $\sum_{i=1}^n \bar \sigma_{ij}\sigma_{ik}=\delta_{jk}$. 
$G_n$ is called orthogonal, unitary or symplectic group accordig as scalars are $\bf{ R,C}$ or $\bf H$\cite{KR90a, Steenrod51}. 
Quarternions $ {\bf H}$ are related to Symplectic transformations, Complex numbers ${\bf C}$ are related to unitary transformations and Real numbers ${\bf R}$ are related to orthogonal transformations.

Properties of $\bf H$ can be studied in the general framework of symplectic vector spaces and Clifford algebra\cite{Souriau70, Sniatycki80, Garling11}.
Artin\cite{Artin88} considered a generalized quaternion algebra in the framework of Clifford algebra and quadratic forms. 

Usefulness of ${\bf H}$ in representing Lorentz transformations of the spinors was presented by Dirac\cite{Dirac45}. Algebra using ${\bf R, C, H}$ is called K-theory and used in noncommutative geometry \cite{Connes90,Connes94,Connes95,Atiyah00}, and applied in the standard model of elementary particles .  

 The Hopf lemma, which guaranties that ``a commutative real algebra without divisor of zero, having quadratic mappings and every element is a square, the algebra is 2-dimensional'', induced complex analytical manifold which is diffeomorphic to tensor product of a circle ${\bf S}^1$ and a sphere ${\bf S}^3$. Rotations of ${\bf S}^3$ are expressed by $\bf H$, and Kodaira\cite{Kodaira92} clarified the structure of the Hopf manifold using complex projective space ${\bf P}^3({\bf C})$, and Hopf algebra was used in quantum groups.

The structure of this article is as follows.
In section 2, quaternions as bases of symplectic vector space is presented,  and in section 3 formulation of Artin on generalized quaternions as bases of Clifford algebra is presented. 
In physical application, orthogonal groups are more familiar than symplectic groups. In section 4, Milnor's formulation \cite{Milnor68} and Grassmann manifolds over quaternion fields are discussed.
In section 5, application of quaternion in parallel transformations in holonomy group and Kodaira's Hopf space analysis on complex projective manifold \cite{Kodaira92} is presented. 
In section 6, Serre's formulation\cite{Serre71} of representing tensor products of $\bf H$ and $\bf C$ as $SL(2,{\bf F}_3)$, where ${\bf F}_3$ is free group on 3 elements, and its application to quantum group based on Hopf algebra are shown. 
Summary and discussion are given in section 7.

\section{\protect\centering{\sc Quaternions and Symplectic Vector Space}}
In classical mechanics, equation of motion is
\[
\frac{d{\bf v}}{d t}=\frac{\bf F}{m}, \quad \frac{d{\bf r}}{dt}={\bf v}
\] 
and when ${\bf F}$ is a derivative of a potential $w$, the Lagrangian $L=\frac{1}{2}m{\bf v}^2-w$ is considered, and
between arbitrary times $t_0$ and $t_1$, $\int_{t_0}^{t_1}L dt$ is called Hamiltonian action\cite{Souriau70}.
When one defines a triplet $y=^t( t,{\bf r},{\bf v})$ and 
\[
dy=\left(\begin{array}{c}
dt\\
d{\bf r}\\
d{\bf v}\end{array}\right),
\delta y=\left( \begin{array}{c}
\delta t\\
\delta{\bf r}\\
\delta{\bf v}\end{array}\right),
\]
a Lagrangian form $\sigma(dy)(\delta y)$ is defined as
\[
\langle m d{\bf v}-{\bf F}dt, \delta {\bf r}-{\bf v}\delta t\rangle-\langle m\delta {\bf v}-{\bf F}\delta t, d{\bf r}-{\bf v} dt\rangle
\]
where $\langle,\rangle$ is the scalar product.
The vector $dy$ is tangent to a leaf $x$ which passes $y$ if
\[
\sigma(dy)(\delta y)=0 \quad \forall \delta y
\]
The direction of the leaf is the kernel of the form $\sigma$.

Sympleqtic vector space is a real vector space $E$ of $n$ dimension on which 2-form $\sigma$ with following properties are defined.

For two vectors $X$ and $Y$ of $E$, when $\sigma(X)(Y)$ is zero, $X$ and $Y$ are called orthogonal. Since $\sigma(Y)(X)=-\sigma(X)(Y)$, the relation is symmetric.
The space $orth(H)$ is defined as a set of vectors $X$ which are orthogonal to all $Y\in H$:
\[
X\in orth(H) \Leftrightarrow \sigma(X)(Y)=0, \quad \forall Y\in H
\]

A vector subspace$H$ is called isotropic if $H\subset orth (H)$, and co-isotrope if $orth(H)\subset H$, respectively. If $H=orth(H)$, the vector space $H$ is called self-orthogonal.
For a base $S=[U_1,\cdots,U_p, V_1\cdots V_p]$ when
\[
\sigma(S(x))(S(y))=\bar x\cdot J\cdot y, \quad x,y\in R^{2p}
\]
where
\[
J=\left ( 
\begin{array}{ccccccc}
& & &\vdots&1 & &\\
& 0 & &\vdots &  &\ddots & \\
& & &\vdots & & & 1\\
\ldots &\ldots &\ldots & \vdots &\ldots &\ldots &\ldots\\
-1 & & &\vdots & & &\\
 & \ddots & &\vdots& & 0 &\\
 & & -1&\vdots & & &
 \end{array}
 \right )
\]
is satisfied, $E$ is called symplectique of dimension $2p$. An element of general linear transformation $A$ that satisfy
\[
\sigma(A(X))(A(Y))=\sigma(X)(Y)
\]
is called symplectic group $Sp(E)$.

For a $p$ dimensional vector field $E$, a manifold $V=E^*\times E$ is defined by
\[
y\equiv \left (\begin{array}{c}
P\\
Q\end{array}\right )\quad P\in E^*,  Q\in E. 
\]
When the Cartan form $\varpi(dy)\equiv P(d Q)$, is a potential, i.e. $\sigma=\nabla\varpi$,  
\[
\sigma(dy)(\delta y)=[dP](\delta Q)-[\delta P](d Q).
\]

For $V'\subset V$, an induced form $\sigma_{V'}$ from the Lagrangian $\sigma_V$ satisfy for $dy,\delta y\in H$
\[
\sigma_{V'}(dy)(\delta y)=\sigma_V(dy)(\delta y)
\]
\[
ker(\sigma_{V'})\equiv H\cap orth(H).
\]

Souriau \cite{Souriau70} defined a torus $T\ni z$, that defines $z=exp(is)$ $(s\in {\bf R}$, $mod  2\pi$), and the quantum manifold $Y$ whose structure is defind by $\xi\to \varpi$. The exterior derivative $\nabla\varpi$ and $\varpi$ satisfy
\begin{eqnarray}
dim(ker\nabla\varpi))&\equiv&1\nonumber\\
dim(ker\varpi\cap ker\nabla\varpi)&\equiv&0\nonumber
\end{eqnarray}

In symplectic space, $\varpi$ is 1-form associated with derivative operations $P$ on $Y$, and $\sigma=\nabla\varpi$ is a 2-form.
One introduces the vector $i_Y(\xi)$ which satisfies
\[
\sigma(i_Y(\xi))\equiv0, \quad \varpi(i_Y(\xi))=1.
\]
When $z=e^{\sqrt{-1}s}\in T$, a Lie group operation on $z$ in the manifold $Y$ is 
$\underline{z}_Y=exp(s i_Y)$. Lagrangian $\sigma$ is defined by the relation
\[
\sigma(dx)(\delta x)\equiv(d\xi)(\delta\xi)
\]
$ker(D(P)(\xi)$ for all $\xi\in Y$ is produced by $i_Y(\xi)$,

When $x\in U$, $P^{-1}(x)$ is an orbit of $T$, and $\sigma_Y(d\xi)(\delta\xi)=\sigma_U(dx)(\delta x)$

When $U$ is quantizable, there exists a differential application $x\to z_{jk}$ such that
\[
\varpi_k(dx)-\varpi_j(dx)\equiv\frac{dz_{jk}}{\sqrt{-1} z_{jk}}
\]
vector field $Z_V$ on manifolds $V$ on which a group $G$ operate as
\[
Z_V(x)=D(\hat x)(e) (Z)
\]
where for $a=e\cdot da=Z\cdot dx=0$, $Z_V(x)=d[a_V(x)].$

Moment $\mu$ of the group $G$ is on $Y$, the form $\varpi$,
\[
\mu\cdot Z=\varpi(Z_Y(\xi))\quad Z\in\mathcal G, \quad \delta_{\mu\cdot Z}\xi=Z_Y(\xi),
\] 
where $\xi=\left(\begin{array}{c}
 z_1\\
 z_2\end{array}\right)$ and  $\bar\xi=(\bar z_1,\bar z_2)$ where $z_1, z_2\in {\bf C}$.
For all elements $a$ and $x$ in the $E$
\[
a_E(x)=a+x, \quad Z_E(x)=Z.
\]
$E$ has the quantification $(Y,P)$:

\begin{itemize}
\item $Y=E\times T$

\item $\varpi \delta\left (\begin{array}{c}
x\\
z\end{array}\right )\equiv\frac{\delta z}{\sqrt{-1} z}+\frac{1}{2}\sigma(x)(\delta x) \quad [z\in T, x\in E]$

\item $P\left ( \begin{array}{c}
x\\
z\end{array}\right )\equiv x $
\end{itemize}

If $Y$ has the Lie group structure, the product has the Weyl group structure\cite{Weyl28}
\[
\left(\begin{array}{c}
x\nonumber\\
z\end{array}\right)\times 
\left(\begin{array}{c}
x'\nonumber\\
z'\end{array}\right )=
\left(\begin{array}{c}
x+x'\nonumber\\
zz' e^{-\sqrt{-1}\sigma(x)(x')/2}\end{array}\right ).
\]
The real one 
$\varpi(\delta\xi)={\bar\xi \delta\xi}/{\sqrt{-1}}$
and when $x\in {\bf S}^2$ and $Z\in {\bf R}^3$
\[
\mu(j(Z))=-\lambda\langle x,Z\rangle
\]

\[
\gamma(\Phi(a)\cdot Z)=a\cdot \gamma(Z)\cdot \bar a
\]
When $G=SO(3)$, it is possible to choose
\[
\sigma(dx)(\delta x)\equiv \lambda \langle x, dx\times \delta x\rangle
\]
and define 
\[
j(z)(y)=z\times y \quad \forall y\in {\bf R}^3
\]
and define the moment $\mu\equiv g(-\lambda x)j^{-1}$.

When $\lambda=1/2$,
\[
j(Z)_Y(\xi)=-\frac{\sqrt{-1}}{2}\gamma(Z)\xi\quad \forall \xi\in Y, \quad \gamma(Z)=\sum_{i=1}^3\sigma_j Z^j
\]
where $\sigma_i$ are $SL(2, {\bf C})$ Pauli matrices
\[
\sigma_1=\left(\begin{array}{cc}
0 &\sqrt{-1}  \\
\sqrt{-1} & 0 \end{array}\right),\quad
\sigma_2=\left(\begin{array}{cc}
0 & -1\\
1 & 0 \end{array}\right),\quad
\sigma_3=\left(\begin{array}{cc}
\sqrt{-1} & 0 \\
0 &\sqrt{-1} \end{array}\right).
\]
which is the rotation on ${\bf S}^2$ expressed by quaternions $\gamma(Z)=\sum_{i=1}^3\sigma_j  Z^j$.

Quaternions $\bf H$ are the orthogonal bases of the Hilbert space\cite{Souriau70, Garling11}, consisting of elements ${\bf I}, {\bf i}, {\bf j}, {\bf k}$ which satisfy 
\begin{eqnarray}
&&{\bf i j}={\bf k},\quad {\bf j k}={\bf i},\quad {\bf k i}={\bf j}\nonumber\\
&&{\bf j i}=-{\bf k},\quad {\bf k j}=-{\bf i},\quad {\bf i k}=-{\bf j}\nonumber.
\end{eqnarray}
\[
{\bf I}=\left(\begin{array}{cc}
1  & 0 \\
0 & 1\end{array}\right),\quad
{\bf i}=\sigma_1\quad
{\bf j}=\sigma_2,\quad
{\bf k}=\sigma_3
\]
and 
\[
{\bf H}={\bf I}a+{\bf i}b+{\bf j}c+{\bf k}d=\left(\begin{array}{cc}
a+\sqrt{-1}d  & -c+\sqrt{-1}b \\
c+\sqrt{-1}b & a-\sqrt{-1}d \end{array}\right),\quad a,b,c,d\in{\bf R}
\]
and
\[
{\bf H}^*={\bf I}a-{\bf i}b-{\bf j}c-{\bf k}d.
\]
We define $a+\sqrt{-1}d=\alpha$ and $c+\sqrt{-1}b=\beta$, $\alpha,\beta\in {\bf C}$, and express
\[
{\bf H}=\left(\begin{array}{cc}
\alpha & -\bar \beta\\
\beta &\bar\alpha\end{array}\right), \quad
{\bf H}^*=\left(\begin{array}{cc}
\bar\alpha &  -\beta\\
\bar\beta &\alpha\end{array}\right).
\]
The subset ${\bf i}b+{\bf j}c+{\bf k}d$ is called $Pu({\bf H})$.

Quaternion appeared in symplectic vector space in the representation of $SO(3)$ rotation group of a point $\bf x$ on ${\bf S}^2=SO(3)/SO(2)$.

Instead of the real quaternion algebras, there are complex quaternion algebras\cite{Weyl28} with elements
\[
{\bf x}=\kappa{\bf I}+\lambda {\bf i}/\sqrt{-1}+\mu {\bf j}/\sqrt{-1}+\nu{\bf k}/\sqrt{-1}
\]  
for $\kappa,\lambda,\mu,\nu\in {\bf R}$. 
 In section 5, we extend ${\bf x}\in {\bf S}^2$ to ${\bf x}\in {\bf S}^3=SO(4)/SO(3)\sim O(4)/O(3)$.

\section{\protect\centering{\sc Quaternions and Clifford Algebra}}
Frobenius' theorem says that the only finite-dimensional real division algebra are the real field ${\bf R}$, the complex field ${\bf C}$ and the algebra of Hamilton's quaternions ${\bf H}$. The theorem can be obtained by using the Clifford algebras, which is defined in $d-$dimensional vector field $E$ with orthogonal bases $e_1, e_2,\cdots, e_d$ with quadratic form $q(x)$, for $x\in E$\cite{Artin88}. A quadratic map  $q({\bf x})=B({\bf x,x})$ satisfies
$q(a{\bf x})=a^2 q({\bf x})$ and 
\[
2B({\bf x,y})=q({\bf x+y})-q({\bf x})-q({\bf y}).
\]
An isometric mapping $\tau:V\to V'$, $\tau$ is a linear isomorphism expressed by $GL(V)$ that satisfies
\[
B(\tau({\bf x}),\tau({\bf y}))=B({\bf x,y})\quad \forall {\bf x,y}\in V.
\]

A Clifford mapping $j:(E,q)\to A$, where vector space $E$ has orthogonal bases $(e_1,e_2,\cdots,e_d)$ and algebra $A$ is defined by elements $a_1,a_2,\cdots ,a_d$, is a linear mapping $j(e_i)=a_i$ such that 
\begin{itemize}
\item  $1\notin j(E)$ 
\item  $(j(x))^2=-q(x)1=-q(x)$ for all $x\in E$.
\end{itemize}
 If additionally $j(E)$ generates $A$, then $A$ is called Clifford algebra. 

We define the set 
\[
E^+=\{a\in E; a^2\ge0\} \quad{\rm and}\quad E^-=\{a\in E; a^2\le 0\}
\]
For $a,b\in E^-$, $\beta(a,b)=-(ab+ba)$ is a bilinear mapping of $E^-\times E^-$ into $A$. Since $\beta(a,b)=\frac{1}{2}(a^2+b^2-(a+b)^2)$ $\beta$ takes values in ${\bf R}$. Depending on the univerality of the mapping \cite{Garling11} $A=span(1,E^-)$ is identical to ${\bf C}$ or ${\bf H}$.

Artin\cite{Artin88} defined in isotropic vector space $V$ with orthogonal geometry, and $A$ be any vector which is not contained in the line $\langle N\rangle$. He cosidered $V=\langle N,A\rangle$ i.e. images of $N$ and $A$ fills $V$, and hyperbolic plane $V=\langle N, M\rangle$, each $N$ and $M$ symplectic
\[
N^2=M^2=0, \quad NM=1.
\]
The subspace $U_0=\langle N_1,N_2,\cdots, N_{r-1}\rangle\perp W$ is orthogonal to $N_r$, but does not contain $N_r$. $P_r=\langle N_r,M_r\rangle\subset U_0^*$.

A vector space over the field $k$ of dimension $2^n$ with basis elements $e_S$ for each subset $S$ of $M$ is defined as $C(V)$ in the space $V$. 

When dim$V \leq 4$, in $C(V)$, $A_i$ are orthogonal bases, and a multiplication is denoted by $\circ$.
\begin{eqnarray}
&&(S_1+S_2+S_3)+S_4=S_1+S_2+S_3+S_4\nonumber\\
&&(S_1+S_2+S_3)\cap T=(S_1\cap T)+(S_2\cap T)+(S_3\cap T)\nonumber
\end{eqnarray}
The basis elements $e_S$ and its extension is
\[
e_S\circ e_T=\prod_{s\in S,t\in T}(s,t)\cdot \prod_{i\in S\cap T}A_i^2\cdot e_{S+T},
\]
where $(s,t)$ equals $+1$ for $s\leq t$, and $-1$ for $s>t$.

The associativity is checked by  \cite{Artin88, Lounesto01,Chevalley46}
{\small \[
(e_S\circ e_T)\circ e_R=\prod_{s\in S,t\in T}(s,t)\cdot \prod_{j\in S+T, r\in R}(j,r) \prod_{i\in S\cap T}A_i^2\cdot
\prod_{\lambda\in (S+T)\cap R} A_\lambda^2\cdot e_{S+T+R}
\]}
which becomes
\[
\prod_{s\in S,t\in T}(s,t)\prod_{s\in S.r\in R}(s,r)\prod_{t\in T,r\in R}(t,r).
\]
Product over $A_\nu^2$ is for $\nu$ that appear in more than one of the sets $S, T, R$. 

$C(V)$ has the unit element $e_\phi$, and the vector $A_i$ is identified with the vector $e_{[i]}$, where the set $\{i\}$ contain all the single element $i$. 
\[
A_i\circ A_i=e_{[i]}\circ e_{[i]}=(i,i)A_i^2 e_\phi=A_i^2.
\]
If $i\ne j$ then
\[
(A_i\circ A_j)+(A_j\circ A_i)=(i,j)e_{[i,j]}+(j,i)e_{[i,j]}=0
\]
When $r=3$, $e_S=A_{i_1}\circ A_{i_2}\circ A_{i_3}$, where $A_1,A_2, A_3$ is an orthogonal basis of $V$,
\begin{eqnarray}
(i_1\circ i_2)&=&-(i_2\circ i_1)=-A_3^2 i_3, \quad i_1^2=-A_2^2 A_3^2,\nonumber\\
(i_2\circ i_3)&=&-(i_3\circ i_2)=-A_1^2 i_1, \quad i_2^2=-A_3^2 A_1^2,\nonumber\\
(i_3\circ i_1)&=&-(i_1\circ i_3)=-A_2^2 i_2, \quad i_3^2=-A_1^2 A_2^2.\nonumber
\end{eqnarray}
Algebra of this form is called generalized quaternion algebra.

 When 
\[
 {A_1}^2=1, \quad {A_2}^2=-1,\quad {A_3}^2=a\in { scalar\, norm}\, k^*
\]
one has
\begin{eqnarray}
(i_1\circ i_2)&=&-(i_2\circ i_1)=-a i_3, \quad i_1^2=+a\nonumber\\
(i_2\circ i_3)&=&-(i_3\circ i_2)=-a i_1, \quad i_2^2=-a\nonumber\\
(i_3\circ i_1)&=&-(i_1\circ i_3)=-a i_2, \quad i_3^2=+1\nonumber
\end{eqnarray}
The multiplication rule can be represented by $2\times 2$ matrices
\[
 1=\left(\begin{array}{cc}
1  & 0 \\
0 & 1\end{array}\right),\quad
i_1=\left(\begin{array}{cc}
0 & a \\
1 & 0 \end{array}\right),\quad
i_2=\left(\begin{array}{cc}
0 & a\\
-1 & 0 \end{array}\right),\quad
i_3=\left(\begin{array}{cc}
1 & 0 \\
0 & -1\end{array}\right).
\]
The space $C(V)$ consists of $C^+(V)$ and $C^-(V)$ of dimension 4. 
 $C^+(V)$ is spanned by even number of quaternions $e_S=A_{i1}\circ A_{i2}$ or $A_{i1}\circ A_{i2}\circ A_{i3}\circ A_{i4}$, and $C^-(V)$ is spanned by odd number of quaternions $e_S=A_{i1}$ or $A_{i1}\circ A_{i2}\circ A_{i3}$.

 $e_S\circ e_S=\prod_{s_1,s_2\in S}(s_1, s_2)\cdot \prod_{i\in S} A_i^2 \in k^*$ , where $(s_1, s_2)$ is a sign which is $+1$ if $s_1\leq s_2$ and $-1$ if $s_1> s_2$, and $A_i^2$ is the ordinary square of the basis of a vector $A$.

For a given $S\ne\phi,M$, there is centralizer, that satisfy for $\alpha=\sum_{S}\gamma_S e_S$, \\
$e_T\circ \alpha\circ e_T^{-1}=\alpha$,
\[
C_0(V)=k+k e_M
\]

Pure quaternions are related to three dimensional even algebra ${Cl_3}^+$
\[
{\bf i}\equiv -e_{23}, \quad {\bf j}\equiv -e_{31},\quad {\bf k}\equiv -e_{12}.
\]
The element $e_{123}$ which corresponds to the Pauli $SL(2,{\bf C})$ matrix form
\[
\left(\begin{array}{cc}
\sqrt{-1} & 0\\
0 & \sqrt{-1}\end{array}\right)
\]
commutes with $\sigma_1,\sigma_2,\sigma_3$ and is called the center element. Center elements in ${\bf H}$ are
\[
\{ w\in {\bf H}|w q= q w\quad \forall q\in{\bf H}\}.
\]
In terms of $Cl_{0,3}\simeq {\bf H}\times {\bf H}$, for ${\bf a,b}\in{\bf R}^{0.3}\subset Cl_{0,3}$
\[
{\bf a}\times {\bf b}=\langle {\bf ab}(1-e_{123})\rangle_1.
\]
 When ${\bf a,b}\in {\bf R}^3\subset {\bf H}$, ${\bf a}\times {\bf b}$ is imaginary part of ${\bf a b}$, and when  ${\bf a,b}\in {\bf R}^3\subset Cl_3$, $-\langle {\bf a b} e_{123}\rangle_1$ 

\pagebreak
 
Euclidean vector space ${\bf R}^7$ can be defined by orthnormal bases $e_1,e_2,\cdots,e_7$ with  antisymmetry
$e_i\times e_j=-e_j\times e_i$, and $e_i\times e_{i+1}=e_{i+3}$, where indices are permuted cyclically and translated modulo 7\cite{Lounesto01}. 
Product of two vectors ${\bf a,b}$ in ${\bf R}^7$ defines ${\bf a}\times {\bf b}=\frac{1}{2}({\bf a}\circ{\bf b}-{\bf b}\circ{\bf a})$ which are called Octonions ${\bf O} ={\bf R}\oplus {\bf R}^7$. 

By using a new imaginary unit $l$ ($l^2=-1$), octonion can be constructed from quaternion by the Cayley-Dickson doubling process\cite{Lounesto01}, ${\bf O}={\bf H}\oplus{\bf H}l$, in analogy to ${\bf H}={\bf C}\oplus {\bf C}j$.

In Clifford algebra, products of octonions $a=\alpha+{\bf a}\in {\bf R}\oplus{\bf R}^7$ and $b=\beta+{\bf b}\in {\bf R}\oplus{\bf R}^7$ is
 \[
 a\circ b=\alpha \beta+\alpha {\bf b}+{\bf a}\beta+{\bf a}\cdot{\bf b}+{\bf a}\times{\bf b} 
 \]
and in ${\bf R}\oplus{\bf R}^{0,7}$
\[
a\circ b=\langle ab(1-{\bf v})\rangle_{0,1}
\]
where ${\bf v}=e_{124}+e_{235}+e_{346}+e_{457}+e_{561}+e_{672}+e_{713}$.

In the expression of ${\bf R}\oplus{\bf R}^{0,7}$, physical parametrization of the scalar part (time of the system) can be chosen uniquely.

\section{\protect\centering{\sc Grassmann Manifolds over Quaternion Fields}}
For a complex $n$ dimensional space $V$, Complex Grassmanian $G_k(V)$ is the set of subspaces of complex codimension $k$ in $V$, which is sometimes called $(n-k)$-plane in $V$\cite{BT82}. It is represented as
\[
G_k(V)=\frac{U(n)}{U(k)\times U(n-k)}
\] 
For a real $n$ dimensional space $V$, real Grassmanian $G_k(V)$ is the set of subspaces of real codimension $k$ in $V$. It is represented as
\[
G_k(V)=\frac{O(n)}{O(k)\times O(n-k).}
\] 
Complex structure on ${\bf R}^n$ defines a linear transformation $J$ that satisfies $J^2=-I$. Geodesics of curves from $I$ to $-I$ on Orthogonal group ${\bf O}(n)$ is homeomorphic to complex structure space $\Omega_1(n)$. \cite{Milnor68}. 
\begin{itemize}
\item $\Omega_1(n)=\Omega {\bf O}(n)$ is sets of complex structures.
\item $\Omega_2(n)$ is sets of quaternion strutures on ${\bf C}^{n/2}$ or the vector space of ${\bf H}^{n/4}$
\item $\Omega_3(16r)$ is all subsets of ${\bf H}^{4r}$, or Grassman manifolds over quaternion fields. 
When $V=V_1+V_2$, $dim_{\bf H} V_1=dim_{\bf H} V_2=2r$.
\item $\Omega_4(16r)$ is sets of isometric operators from $V_1$ to $V_2$. It is isomorphic to symplectic group ${\bf Sp}(2r)$
\item $\Omega_5(16r)$ is sets of vector space $W\in V_1$ such that 1) W is closed in $J_1$ and 2) $V_1=W_1\oplus J_2W$.
\item $\Omega_6(16r)$ is sets of real subsets $X\in W$, such that $W$ can be decomposed as $X\oplus J_1 X$.
\item $\Omega_7(16r)$ is sets of all real Grassmann manifolds of $X\simeq {\bf R}^{2r}$.
\item $\Omega_8(16r)$ is sets of all real isometric operators from $X_1$ to $X_2$.
\end{itemize}
Bott showed that homotopy group $\pi_i{\bf O}$ is isomorphic to $\pi_{i+8}{\bf O}$

 van Baal et al \cite{vanBaal92,vBC92,vBHD92,vBvdH94} defined $\bar \sigma_\mu=({\bf I}, \sqrt{-1}{\bf i}, \sqrt{-1}{\bf j},\sqrt{-1}{\bf k})$, \\
 $\sigma_\mu=({\bf I}, -\sqrt{-1}{\bf i}, -\sqrt{-1}{\bf j},-\sqrt{-1}{\bf k})$ whose products are expressed by 
\[
\sigma_\mu\bar\sigma_\nu=\eta^{\mu\nu}_\alpha \bar\sigma_\alpha, \quad \bar\sigma_\mu\sigma_\nu=\bar\eta^{\mu\nu}_\alpha \sigma_\alpha,
\]
where $\eta^{\mu\nu}_\alpha=\bar\eta^{\mu\nu}_\alpha$ are real 't Hooft symbol\cite{tHooft76}, 
\begin{eqnarray}
\eta^{\mu\nu}_\alpha&=&\epsilon_{\mu\nu a},  \quad \eta^{4\nu}_\alpha=-\delta^\nu_\alpha \quad (\mu,\nu,\alpha=1,2,3),\nonumber\\
\eta^{\mu 4}_\alpha&=&\delta^\mu_\alpha,\quad \quad (\mu,\alpha=1,2,3), \quad \eta^{44}_\alpha=0 \nonumber
\end{eqnarray}
and $Q=q_\mu\bar\sigma_\mu$ are quaternions.

Dirac\cite{Dirac45} applied quaternions to Lorentz transformations. He considered a quaternion
\[
{\bf q}={\bf I}q_0+{\bf i}q_1+{\bf j}q_2+{\bf k}q_3
\]
and a vector $A_\mu$ in space-time, whose square is

\pagebreak

\[
A_0^2-A_1^2-A_2^2-A_3^2.
\]
$q$ is expressed as the ratio of quaternions $q=u v^{-1}$, and Lorentz group was defined by introducing three quantities
\[
Q_1=u\bar v, \quad Q_2=u\bar u, \quad Q_3=v\bar v
\]
When $u$ and $v$ are replaced by $\lambda u$ and $\lambda v$, 
\begin{eqnarray}
u\lambda(\bar{\lambda v})&=&u\lambda\bar \lambda\bar v=Q_1\lambda\bar \lambda\nonumber\\
Q_1&=&X_0+{\bf i}X_1+{\bf j}X_2+{\bf k}X_3
\end{eqnarray}
$Q_2$ and $Q_3$ were put
\[
Q_2=X_4-X_0, \quad Q_3=X_4+X_5
\]
Dirac showed that $\xi_\nu=X_\nu/X_0$ and $\eta_\nu=X_\nu/X_5$ ($\nu=1,2,3,4$) transform like space-time vectors. For $f=(1/2)(1+{\bf I})$, the spinorial basis is defined by $f X_0$ and $f X_5$ \cite{Vaz97}.

Lorentz transformation was performed by
\[
q^*=(a q\pm \mu a)(-\mu a q\pm a)^{-1}
\]
where $a$ is an arbitrary quaternion and $\mu$ is a pure imaginary quaternion.

Quaternions represent rotation of 3-dimensional sphere ${\bf S}^3$.
The manifold ${\bf S}^3\times {\bf R}$, where ${\bf R}$ represents periodic and anti-periodic time variables is ihomeomorphic to $T^4$. Applications of quaternions as rotation operator are discussed in \cite{Kuipers99,Hughes00, RRPJ05}. 
We see in the next section that the manifold of ${\bf S}^1\times {\bf S}^3$ has different Riemann surface structure than that of $T^4$.

\section{\protect\centering{\sc Holonomy Group and Quaternion}}
When self-orthogonal subspace of $E$ is $H$ and the Lagrangean form is defined by $\sigma$, $ker(\sigma)\subset H$, the base of $ker(\sigma)$ is written as $T$ and the base of $H$ is written as $[T,U]$. In this case
the transformation matrix considered in section 2, $S=[U_1,\cdots, U_p,V_1,\cdots,V_p]$ changes to  
$S=[T_1,\cdots, T_q, U_1,\cdots, U_p, V_1,\cdots,V_p]$ and the matrix of $\sigma$ components becomes

\[
J=\left ( 
\begin{array}{cccccccccccc}
&&&\vdots&&&&\vdots&&&\\
& 0 & & \vdots&& 0 & &\vdots && 0 &\\
&&&\vdots&&&&\vdots&&\\
\ldots&\ldots&\ldots&\vdots&\ldots &\ldots &\ldots & \vdots &\ldots &\ldots &\ldots\\
&&&\vdots&& & &\vdots&1 & &\\
&0&&\vdots&& 0 & &\vdots &  &\ddots & \\
&&&\vdots&& & &\vdots & & & 1\\
\ldots&\ldots&\ldots&\vdots&\ldots &\ldots &\ldots & \vdots &\ldots &\ldots &\ldots\\
&&&\vdots&-1 & & &\vdots & & &\\
&0&&\vdots& & \ddots & &\vdots& & 0 &\\
&&&\vdots& & & -1&\vdots & & &
 \end{array}
 \right )
\]
where we assume $q=dim(ker(\sigma))$; $q+p=dim(H)$; $q+2p=n$. It is possible to choose another base $S'=[T,U,V']$ of $E$ and matrices $M'_{kj}=\sigma(V'_k)(U_j)$, and $V''=V'\cdot M^{-1}$.

$V''$ is a new base of $V'$ that satisfy 
\[
\sigma(U_j)(V''_k)=\left\{
\begin{array}{ccc}  1&{\rm if}&j=k\\
                                                         0&{\rm if}& j\ne k 
\end{array}\right.
\]
One takes 
\[
V_k=V_k''+\frac{1}{2}\sum_j U_j \sigma(V''_j)(V''_k)\quad k=1,2\cdots p. 
\]
 The space $E$ is called pre-symplectic.

We consider complex manifolds $M={\bf C}^4$, $S=(0,0,0,0)$ and their submanifold without singular points $\bar S={\bf P}^3$, which is complex projective space covered by $u_j\in {\bf C}^3: {\bf P}^3=\sum_{j=0}^3 u_j$.
Points $\zeta$ on ${\bf P}^3$ and $t$ on ${\bf R}$ define maps 
\[
\Phi :(t,\zeta_1.\zeta_2)\to  (z_1,z_2)=(\zeta_1 e^{t\beta_1}, \zeta_2 e^{f\beta_2}),
\]
 which is 1 to 1 mapping from ${\bf R}\times {\bf S}^3\to W$ defined by $z_1, z_2$.
 For maps $C^*: (\zeta_1,\zeta_2,\zeta_3,\zeta_4)\to (g\zeta_1,g\zeta_2, g\zeta_3, g\zeta_4)$, $W/C^*={\bf P}^3$
 and  $g^m: (t,\zeta_1,\zeta_2)\to (t+m,\zeta_1,\zeta_2) , m\in {\bf Z}$.
 Therefore $W/G={\bf R}/{\bf Z}\times {\bf S}^3={\bf S}^1\times {\bf S}^3$ \cite{Kodaira92}.
  
A Lagrangian system with linear constraints are holonomic or non-holonomic \cite{Arnold78} if imposed constraints are integrable or not. Condition is characterized by the parallel transformation of the solution curves in complex plane or quaternion tangent plane.

Consider a group $G_n$ which is transitive on the unit sphere $S$, there is the map 
\[
p_1: \quad G_n\to S 
\]
Let $M_n$ be the projective space associated with vector fields $Q_n$.  When non-zero elements of $Q_n$, $x$ and $y$ have relation $y=x q$, ($q\in Q$), they belong to equivalence classes, and there is a projection
\[
p: \quad S\to M_n
\]

 If $H$ is the subgroup of $G_n$ which has a fixed point $z_0=p(x_0)$ and there is a map 
 \[
 p_2 : \quad G_n\to M_n, \quad M_n=G_n/H,
 \]
then $H$ can be expressed by the direct product $H=Q'\times G_{n-1}$, where $Q'$ is the ${\bf S}^3$, and vector fields consist of quaternions of $|q|=1$ \cite{Steenrod51}.

Choosing a complex analytic space $W={\bf C}^2-\{0\}$ and mappings
\[
g: z(z_1,z_2)\to g(z)=(\alpha_1 z_1 \alpha_2 z_2)
\]
and $G=g^m|_{m\in {\bf Z}}$, where $|\alpha_1|>1, |\alpha_2|>1$ are constants. The manifold $W/G$ is called Hopf manifold, which is diffeomorphic to ${\bf S}^1\times{\bf S}^3$.

 In the real case ${\bf S}^3\to {\bf R}P^3$ mapping is possible, and analogously in the complex case ${\bf S}^1\to{\bf S}^3\to {\bf C}P^3$
is possible\cite{Hatcher02}.  Since ${\bf C}P^1={\bf S}^2$,  ${\bf S}^3$ is a complete image of ${\bf S}^2$, and the map ${\bf S}^3\to {\bf S}^2$ is called the Hopf map \cite{Steenrod51, KR90b, Hatcher02}.

Kodaira\cite{Kodaira92} and Kodaira and Spencer\cite{KS58} used complex projective line ${\bf P}^1={\bf C}\cup \infty=U_1\cup U_2$, where $U_1={\bf C}$, $U_2={\bf P}^1-\{0\}$,  and defined complex analytical functions in
\[
M_t=U_1\times {\bf P}^1\cup U_2\times {\bf P}^1,\quad t\in {\bf C}.
\]

When $(z_1,\zeta_1)\in U_1\times {\bf P}^1$ and $(z_2,\zeta_2)\in U_2\times {\bf P}^1$ satisfy
\[
z_1\cdot z_2=1, \quad \zeta_1=z_2^2\zeta_2+t z_2
\]
the two points are regarded as identical. 

He used conformal  mappings of 
\[
W_j=\{(z_j,s)||z_j|<1,|s|<1\}=U_j\times D
\]
where $D=\{s\in{\bf C}||s|<1\}$, and considered when the condition 
\[
z_j=f_{jk}(z_k,s)=(f_{jk}^1(z_k,t),f_{jk}^2(z_k,t))
\]
is satisfied $(z_j, s)$ and $(z_k,s)$ are identical in the projected space.

He showed that dimention of cohomology $dim H^1(M_t,\Theta_t)$, where $\Theta_t$ is the sheaf of germs of conformal tangent vector space,  is 1 for $t=0$ and 0 for $t\ne 0$. The discontinuity suggests validity of the difference of pre-symplectic structure of quantum dynaics and symplectic structure of classical dynamics.
General summary of works of Kodaira and Spencer are given in \cite{Hirzebruch78}.

Using $\phi: A\otimes A\to A$ and its dual $\psi$, Hopf algebra $(A,\phi,\psi)$ defines algebra of complex projective spaces.  ${\bf P}^1({\bf C})$ consists of $\zeta=(1,z_1)=(1,\zeta_1/\zeta_0)$, $\zeta_0\ne 0$. Mapping  from ${\bf P}^1({\bf C})$ to a  $3-$dimesional projective plane ${\bf P}^3({\bf C})$: ${\bf P}^1\to {\bf P}^3$ were expressed by Cavalieri and Miles \cite{CM16} by
\[
\phi([S:T])=[S^3:S^2T: S T^2: T^3]\equiv[X:Y:Z:W]
\]
which defines a twisted cubic $V({\bf P})$ in ${\bf P}^3$. 

Let $P_1,P_2$ be homogenious polynomials in 4 variables $P_1=XW-YZ, P_2=XZ-Y^2$, for example, and define
\[
V({\bf P})=\{[X,Y,Z,W]|P_1(X,Y,Z,W)=P_2(X,Y,Z,W)=0\}\subset {\bf P}^3({\bf C}).
\]
For any point $p$ on a three dimensional real sphare ${\bf S}^3\subset {\bf C}^2$, there is a unique complex line ${ l}_p$, or there is a function $H$ such that
\begin{eqnarray}
H: &&{\bf S}^3\to {\bf P}^1({\bf C})\nonumber\\
&& P\to { l}_P. \nonumber
\end{eqnarray}
has the image of Riemann surface isomorphic to ${\bf P}({\bf C})$ which is called a twisted circle in ${\bf P}^3({\bf C})$.

In \cite{CM16}, locus of $P_3=YW-Z^2$ is also considered and the vanishing locus of any two of the three polynomials is strictly larger than the rest polynomial.  

When we replace $\bf C$ to $\bf H$ and consider the mapping ${\bf S}^3\to{\bf S}^7\to {\bf S^4}={\bf H}P^1$\cite{Hatcher02}
\begin{eqnarray}
&&{\bf C}^4\setminus\{(0,0,0,0)\}\stackrel{\rm P}{\longrightarrow}{\bf  C}{ j}\nonumber\\
&&\downharpoonleft \pi\nonumber\\
&&{\bf P}^3({\bf C} { j})\nonumber
\end{eqnarray}
where ${\bf C}{ j}$ is $Pu({\bf H})$, we obtain 
\begin{eqnarray}
&&P_1=S^3T^3-S^2TST^2=ST-TS=2ST, \nonumber\\
&&P_2=S^3ST^2-S^2TS^2T=S^2-T^2,\nonumber\\
&&P_3=S^2T T^3-ST^2 ST^2=T^2-S^2, \nonumber
\end{eqnarray}
in which $TS=-ST, S^2=-1$ and $T^2=-1$ are used.
Locus of $P_2$ and $P_3$ give no additional conditions on twisted cubic in ${\bf P}^3({\bf C})$, but that of $P_1$ defines a plane.

Mathematically, structure of complex manifolds of complex dimension 2, ${\bf C}^2\sim {\bf R}^4$ was studied by Russian groups\cite{Rohlin52,Pontryagin87} and Japanese groups.
 Kodaira\cite{Kodaira65,Kodaira92} showed that the ${\bf S}^1\times {\bf S}^3$, where ${\bf S}^1$ represents holonomy group of quaternion variables defined by ${\bf S}^3$, is homeomorphic to Hopf space and have 2 linearly independent vectors.
 
Rotations of ${\bf S}^3$ is expressed by quaternions, whose operations are given by symplectic groups. Quaternions can be expressed by Pauli matrices which follow orthogonal groups $O(3)$, which allowes rotations of ${\bf S}^2$ locally. However the local infinitesimal quantomorphism in \cite{Souriau70} is too restrictive, when one patches conformally transformed curves globally. 
 
Motions of particles in space-time represented by Lagrangian $L(t, {\bf q}, {\bf \dot q})$ are related by holonomy curves, which characterize the structure of the manifold on which curves are defined.
 In case of magnetic interactions the magnetic field ${\bf B}$ whose coordinate is chosen along the $y-$axis, is a function of the magnetizing field ${\bf H}={\bf B}-{\bf M}/{\epsilon_0 c^2}$ whose coordinate is chosen along the $x-$axis, shows hysteresis curves, which indicates that nonholonomic effects appear in Nature.
 
In Hamiltonian formulation of classical mechanics, change of coordinate from $({\bf p}, {\bf q}, t)$, 
\[
z_k=(p_k+i q^k)/\sqrt 2, \quad \bar z_k=(p_k-i q^k)/\sqrt 2
\]
defines a 2-form $\omega=-i \sum_k d\bar z_k\wedge dz_k$.
\begin{figure}[htb]
\begin{center}
\includegraphics[width=6cm,angle=0,clip]{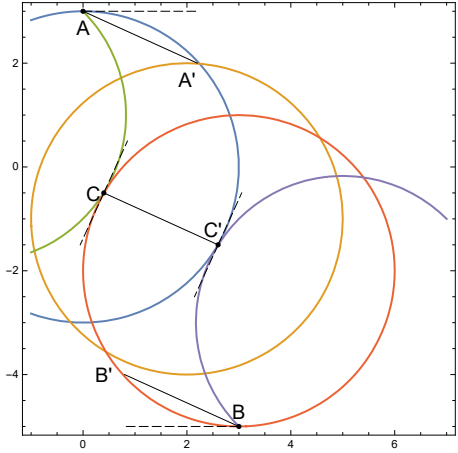} 
%{\small
\caption{ Holonomy curve $A C' B+ B C A$, and $(C B')(B C')+(C'A')(AC)$. The parallel transformation of the former curve have the direction parallel to the $x$ axis, as shown by dashed lines from $A$ and $B$. The parallel transformations of the latter curve have the direction parallel to $CC'$, from $A$ and $B$ respectively, as $AA'$ and $BB'$. The system is holonomic if quaternion vectors $e_x$ and $e_{CC'}$ are parallel. Curves on the manifold of ${\bf S}^1\times {\bf S}^3$ is non-holonomic, since one ${\bf P}({\bf C})$ of ${\bf P}^3({\bf C})$ can be interchanged with ${\bf P}({\bf C})$ of ${\bf S}^1$ in complex projective spaces. Figure copied from the reference \cite{Furui19}. }
%}
\label{holonomy2}
\end{center}
\end{figure}

\begin{figure}[htb]
\begin{center}
\includegraphics[width=7cm,angle=0,clip]{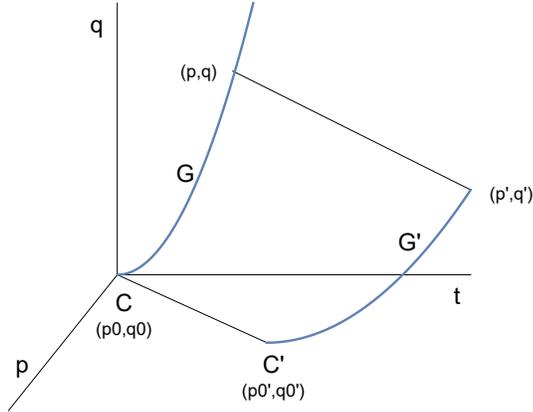} 

\caption{ Lowering the order of a Hamiltonian system. The curve in space-time $G'$ from $(p_0',q_0')$ to $(p',q')$ in pre-symplectic form is projected to the curve $G$ from $(p_0,q_0)$ to $(p,q)$ in symplectic form.  $(p_0,q_0)$ is obtained from $(p_0',q_0')$ by a parallel transformation, and $(p,q)$ is obtained from $(p',q')$ by the parallel transformation of the $Pu({\bf H}): e_{\bf H}$. Actions along $G'$ and along $G$ are not the same, when the system is non-holonomic.}
\end{center}
\end{figure}

The assumption of maximum conformality allows to express the solution by patching local solutions of evolution equations.
 
Extension to $K({\bf P},{\bf Q},t)=H({\bf p},{\bf q}, t)+\frac{\partial S}{\partial t}$ is performed by choosing
\begin{eqnarray}
S({\bf p}_1,{\bf q}_1,t)&=&\int_{({\bf p}_0,{\bf q}_0,t)}^{({\bf p}_1,{\bf q}_1,t)}{\bf p}d{\bf q}-\int_{({\bf p}_0',{\bf q}_0',t)}^{({\bf p}_1',{\bf q}_1',t)}{\bf P}d{\bf Q}\nonumber\\
&=&\int_G dS-\int_{G'}dS\ne 0\nonumber
\end{eqnarray}
when there are hysteresis. In Fig.2 $CC'$ is the parallel transformation by quaternion $e_{CC'}$. On the curve $G$, line-integral $\int _G dS$ becomes 0 due to symplectic structure of the bases, while on the curve $G'$,  although $ddS=0$, the Stokes theorem says for an area $\sigma$ surrounded by $(C B'B C')+(C'A'AC)$ that
\[
\int_{(C B'B C')+(C'A'AC)}{\bf P}d{\bf Q}-H dt=\int_{\partial\sigma}{\bf P}d{\bf Q}-H dt=\int_\sigma d({\bf P}d{\bf Q}-Hdt)
\]
is not necessary 0.

The Hamiltonian $K({\bf P},{\bf Q},t)=H({\bf p},{\bf q},t)$ satisfies the canonical form
\[
\frac{d{\bf P}}{dt}=-\frac{\partial K}{\partial {\bf Q}}, \quad \frac{d{\bf Q}}{dt}=\frac{\partial K}{\partial{\bf P}}.
\]

Arnoldt assumes that the equation $h=H(p_1,\cdots ,p_n,q_1,\cdots, q_n)$ can be solved for $p_1=K({\bf P},{\bf Q},T;h)$, where ${\bf P}=(p_2,\cdots,p_n)$, ${\bf Q}=(q_2,\cdots,q_n)$, $T=-q_1$. Then\cite{Arnold78}
\[
{\bf p}d{\bf q}-H dt={\bf P}d{\bf Q}-K dT-d(H t)+t dH.
\]

%\newpage
For a physically distinguished configuration space $Y$ and Hamiltonian vector fields $\xi_f, \xi_g$,
\[
	[\xi_f,\xi_g]=-\xi_{[f,g]},
\]
and the Lagrange bracket is $\omega=d\theta_Y$. $\theta_Y$ is canonical 1-form\cite{Souriau70, Sniatycki80}.
In terms of local chart $(\pi^{-1}(U),q^1.\cdots,q^4,p_1,\cdots.p_4)$, the symplectic form for a charge $e$ and pullback 2-form $f$, $\omega_e=d\theta_Y+e \pi^* f$ is
\[
\omega_e|_{\pi^{-1}(U)}=\sum_{i=1}^4 dp_i\wedge dq^i+e/2\sum_{ij}(f_{ij}\circ \pi)dq^i\wedge dq^j
\]

\section{\protect\centering{\sc Quaternions in Non-Commutative Geometry and Quantum Groups}}
Serre\cite{Serre71} defined a group $G$ as a union of subgroup of 8 elements
\begin{equation}
E=\pm {\bf I}, \pm {\bf i}, \pm{\bf j}, \pm{\bf k}
\end{equation}
and 16 elements
\begin{equation}
(\pm{\bf I}\pm{\bf i}\pm{\bf j}\pm{\bf k})/2,
\end{equation}
where ${\bf I},{\bf i}, {\bf j}$ and ${\bf k}$ are bases of quaternions. $G$ is called an entire quaternion of Hurwitz\cite{Serre71}. Relations between Hurwitz number and Riemann space are written in \cite{CM16}.

A group $G$ is called resoluble, if there is a series
\[
\{1\}=G_0\subset G_1\subset\cdots\subset G_n=G
\]
and $G_{i-1}$ different from $G_i$ and $G_i/G_{i-1}$ is commutative for $1\leq i\leq n$.
Similarly $G$ is called hyper-resoluble if $G_i/G_{i-1}$ is cyclic.

The group $G$ is invertible with quaternion ring, and is isomorphic to the special linear transformation $SL(2,{\bf F}_3)$, where ${\bf F}_3$ are free group on 3 elements\cite{Serre71}. 
Tensor products of quaternions $\bf H$ and complex numbers ${\bf C}$
\[
{\bf H}\otimes _{\bf R} {\bf C}={\bf M}_2({\bf C})
\]
define the group $G$. 

This construction has an application in quantum groups, which are defined on algebraic Hopf group., 
elements of which are not necessarily commutative / noncommutative\cite{Jimbou90}. A representation of Hopf  algebra $(\pi, V)$ consists of a vector space $V$ and algebraic endomorphism $\pi: A\to End(V)$. The algebra $A$ contains coproduct
\[
\Delta: A\to A\otimes A
\]
and coalgebra morphism 
\[
\epsilon: A\to {\bf C}.
\].

Universal enveloping algebra $U_q(sl(2,{\bf C}))$ of Jimbou\cite{Jimbou90} and Drinfeld\cite{Drinfeld85} contains generators $X^+, X^-, K, K^{-1}$, where $X^+=Max(X,0)$ and $X^-=Max(-X,0)$ are Riesz space variables\cite{Souriau70},  with the following relations
\begin{eqnarray}
K K^{-1}&=&K^{-1} K=1\nonumber\\
K X^\pm K^{-1}&=&q^{\pm 2}X^\pm \nonumber\\
([ X^+, X^- ])&=&\frac{K-K^{-1}}{q-q^{-1}}. \nonumber
\end{eqnarray}
There are algebraic morphism from  $X^+, X^-, K, K^{-1}$ of  $U_q(sl(2,{\bf C}))$ to $X_i^+, X_i^-, K_i, K_i^{-1}$ $(1\leq i\leq l)$ of  $U_q(\hat{sl}(2,{\bf C}))$ that satisfy \cite{Jimbou90}
\begin{eqnarray}
\phi({X_0}^\pm)&=&X^{\mp}, \quad \phi({X_1}^\pm)=X^\pm\nonumber\\
\phi(K_0)&=&K^{-1}, \quad \phi(K_1)=K\nonumber
\end{eqnarray}
\begin{eqnarray}
K_i K_i^{-1}&=&K_i^{-1}K_i=1,\quad K_i K_j=K_j K_i\nonumber\\
K_i {X_j}^{\pm}{K_i}^{-1}&=&{q_i}^{\pm a_{ij}}{X_j}^{\pm}\nonumber\\
([{X_i}^+, {X_j}^- ])&=&\delta_{ij}\frac{K_i-{K_i}^{-1}}{q_i-{q_i}^{-1}}\nonumber
\end{eqnarray}
where $a_{ij}$ is the element of the General Cartan Matrix. Derivation of $U_q({\mathcal G})$ from $U_q(\hat{\mathcal G})$ is satisfied for $a_{ij}$ of $\hat{\mathcal G}\in A_l$ type Lie group\cite{Jimbou90}.

One can construct  $SL(2,{\bf F}_3)$ from $\{K_0,K_1, X_1^\pm\}$ using quaternions, but $\phi$ does not preserve coproducts. One can define automorphism $T_\lambda$ ($\lambda\in {\bf C}$) such that
\[
T_\lambda({X_0}^\pm)=\lambda^\pm X_0^\pm, \quad T_\lambda(X)=X \quad(X=X_1^\pm,K_0,K_1)
\]
and derive representations of $U_q(sl(2,{\bf C}))$ depending on the parameter $\lambda$\cite{Jimbou90}.
Hence quaternions define a morphism of Hopf algebra which is dependent on a parameter $\lambda$.

Differential calculus on the real quantum plane and mapping of quantum groups on non-commutative differential manifolds did not show qualitative difference of Stokes' theorem from that of classical groups  \cite{WZ90}. An extension to complex quantum plane was discussed in \cite{PW90}. 

\section{\protect\centering{\sc Summary and Discussion}}
We started from 2-form in symplectic space and from invariance of Lagrangian 2-forms, derived linear combination of vectors with Pauli $SL(2,{\bf C})$ matrices as bases and obtained pure quaternions. From Clifford algebra with quadratic forms, we obtained generalized quaternions, and found that in addition to real numbers ${\bf R}$ and complex numbers ${\bf C}$, quaternions ${\bf H}$ form finite division algebra.
 
In variational analysis of dynamics, centralizer of dynamical transformations is important, and dynamics follows the $SO(3)$ transformation, parallel transformations by quaternion vectors play important role in non-holonomic systems.  In standard holonomic systems, parallel transformations by real vectors are considered.

Guillou and Marin\cite{GM86} explained the work of Rohlin\cite{Rohlin52} and Pontryagin\cite{Pontryagin87}.  Pontryagin showed in mappings of $(n+2)-$dimensional sphere $\Sigma^{n+2}$ into an $n-$dimensional sphere ${\bf S}^n$, there are exactly two classes of mappings, when $n\ge 2$.

The theorem can be applied to the existence of holonomy curves $\alpha:$ $ACB\oplus BC'A$ and $\beta:$ $ACB'\oplus BC'A'$, shown in  the Figure 1. ($B$ and $B'$ are identified by the parallel transformation and $A$ and $A'$ are identified similarly.)  The index of intersection \cite{Pontryagin87} $J(\alpha_i,\alpha_j)=J(\beta_i,\beta_j)=0$, but $J(\alpha_i,\beta_j)=\delta_{ij}$ where $i,j$ for the curve $\alpha$ indicates indices of the set $\{AC, CB, BC', C'A\}$ and for the curve $\beta$ indicates indices of the set $\{AC, CB', BC', C'A'\}$. 
The pre-symplectic structure of the vector space $E$, which allows presence of different kernel space of parallel transformations is essential for the quantum dynamics, while classical dynamics can be understood by symplectic structure of the vector space.

On the manifold of ${\bf S}^1\times {\bf S}^3$, we consider exactly two mappings from the sphere $\Sigma^5$ into ${\bf S}^3$\cite{Pontryagin87}, where ${\bf S}^3$ is defined by complex variables
\[
(z_1, z_2 |z_1\bar z_1+z_2\bar z_2=1)
\]
which has the same number of degrees of freedom as that of quaternions,  and ${\bf S}^1$ of quaternions defines a group ${\bf g}\in G$, satisfying $\psi_g({\bf x})={\bf g} {\bf x} {\bf g}^{-1}$ is chosen to satisfy $\psi_g({\bf i})={\bf i}$, or ${\bf g}=\cos\theta+{\bf i}\sin\theta$.

The pre-symplectic form vector space is important in propagation of solitons in hysteretic media\cite{FdS19}. When one discretizes time in evolution equations, one obtains recursion operator. In the case of viscous Burger's equation\cite{Hoermander97}
\[
u_t+u\partial u/\partial x=\mu \partial^2 u/\partial x^2; \quad t\ge 0; \quad u(0,x)=u_0(x); \quad \mu>0,
\]
the equation can be written as
\[
\partial u/\partial t+\partial f(u)/\partial x=0,\quad u(0,x)=u_0(x),
\]
whose discretized form is
\[
u_n^{k+1}-(u_{n+1}^k+u_{n-1}^k)/2)/h+(f(u_{n+1}^k)-f(u_{n-1}^k))/2l=0
\]
where grid points of half plane $\bar{\bf R}_+\times {\bf R}$ are $\{(kh,nl); k,n\in {\bf Z},k\ge 0\}$. The first term contains a term
\[
(2u_n^k-u_{n+1}^k-u_{n-1}^k)/2h.
\]
Discretization in time affects symplectic structure of the vector space of solutions.

Dosch et al\cite{dTB09, DdTB15, BdTDE15} studied spectrum of mesons and baryons in supersymmetric light front QCD embedded in $AdS_5/CFT_4$ space. Conformal algebra alows patching analytical solutions of different times, and the mass spectra of mesons and baryons show specific proportionalities.

Understanding quaternions is relevant to understanding non-commutative geometry, and via Lie algebra, it is related to understanding Heisenberg's hamiltonian and quantum groups.
%\begin{center}
%{\bf Acknowledgement}
%\end{center}
\subsection*{Acknowledgment}
I learned dynamics on symplectic vector spaces and use of quaternions from the book of J. -M. Souriau, which I found in France when I was a boursier of the French Government. Studies in Germany and France after the doctor course were helpful for writing this review.  I thank the organizations of the two countries, and Libraries of Tokyo Institute of Technology and the Library for Mathematical Science of the University of Tokyo for allowing consultation of references.

\renewcommand\refname{\centerline{\sc References}\global\def\bibname{References}}

\bigskip

%\label{lastpage-01}

\begin{thebibliography} {50}
\bibitem{Hamilton} Hamilton, R.W. 1844. {``On quaternions; or on a new system of imaginaries in Algebra''},London, Edinburg, and Dublin Philosophical Magazine and Journal of Science {\bf 25} (169) 489-495.
\bibitem{KR90a} Koecher, M. and Remmert, R. 1990. {``Hamilton's Quaternion'', Chapter 7 of ``{\sl Number}'', 2nd English Edition}, Translated from german original version ``Zahlen'', Springer-Verlag, New York.
\bibitem{Steenrod51} Steenrod, Norman. 1951. {\sl The Topology of fibre Bundles}, Princeton University Press.
\bibitem{Souriau70} Souriau, J. -M. 1970. {\sl Structures des Syst\`emes Dynamiques}, DUNOD, Paris.
\bibitem{Sniatycki80} \'Sniaticki, J. 1980. {\sl Geometric Quantization and Quantum Mechanics}, Springer-Verlag, New York.
\bibitem{Garling11} Garling, D.J.H. 2011. {\sl Clifford Algebras: An Introduction}, London Mathematical Society, Student Texts 78, CambgidgeUniversity Press
\bibitem{Artin88} Artin, Emil 1988. {\sl Geometric Algebra}, Wiley Classics Library, New York. John Wiley \& Sons Inc. pp. x+214

\bibitem{Dirac45} Dirac, P.A.M. 1945. {``Application of Quaternions to Lorentz Transformations''}, Proc. Roy. Irish Acad. (Dublin), {\bf A 50} 261-270.

\bibitem{Connes90} Connes, Alain 1990. {\sl G\'eom\'etrie non commutative} Inter-Editions, Paris, Translated to Japanese by Maruyama, Fumitsuna, Iwanami Shoten Pub. Tokyo.
\bibitem{Connes94} Connes, Alain 1994, {\sl Noncommutative Geometry},Academic Press, An Imprint of Elsevier, San Diego, New York, Boston, London, Sydney, Tokyo, Toronto.
\bibitem{Connes95} Connes, Alain 1995. {``Noncommutative geometry and reality''}, J. Math. Phys.{\bf 36} (11) 6194-6231.
\bibitem{Atiyah00} Atiyah, Michael 2000. {``K-Theory Past and Present''} arXiv:math/0012213v1 [math.KT].
\bibitem{Kodaira92} Kodaira, Kunihiko 1992. {\sl Theory of Complex Manifolds} (In Japanese), Iwanami Shoten Pub.
\bibitem{Milnor68} Milnor, J. 1963. {\sl Morse Theory} 3rd Ed. , Ann. of Math. Studies 51, Princeton University Press; Translated to Japanese by Shiga, Koji 1968. Yoshioka Shoten Pub. Kyoto.
\bibitem{Serre71} Serre, Jean-Pierre 1971. {\sl Repr\'esentations Lin\'eaires des Groupes Finis} Deuxi\`eme \'Edition, Refondue, Hermann, Paris.
\bibitem{Weyl28} Weyl, Hermann 1928. {\sl The Theory of Groups and Quantum Mechanics} Translated from the second German edition by Robertson, H.P. 1951 Dover Publications, inc.
\bibitem{Lounesto01} Lounesto, Pertti 2001. {\sl Clifford Algebras and Spinors} 2nd Edition, Cambridge University Press.
\bibitem{CM16} Cavalieri, Renzo and Miles, Eric 2016. {\sl Riemann Surfaces and Algebraic Curves, A First Course in Hurwitz Theory}, Cambridge University Press. p.46
\bibitem{Vaz97} Vaz, Jayme Jr. 1997. In {\sl Geometry, Topology and Physics, Proceedings of the First Brazil-USA Workshop held in Campinas, Brazil, June 30-July 7, 1996} Edited by Apanasov, B.N. et al. Walter de Gruyter, Berlin New York, p.277-300.
.
\bibitem{Madore99} Madore, John. 1999. {\sl An Introduction to Noncommutative Differential Geometry and its Physical Applications} 2nd Edition, Cambridge University Press.
\bibitem{Kodaira65} Kodaira, Kunihiko 1965. {``Complex Structures on $S^1\times S^3$''}, Mathematics {\bf 55} 240-243, Proc. N.A.S.

\bibitem{KS58} Kodaira, K. and Spencer, D.C. 1958. {``On the Deformation of Complex Analytic Structures I,II''}, Annals of Mathematics {\bf 67}, p.329, p.403
\bibitem{Hirzebruch78} Hirzebruch, Friedrich 1978. {\sl Topological Methods in Algebraic Geometry} Second, Corrected Printing of the Third Edition, Springer-Verlag, Berlin Heidelberg New York.
\bibitem{BT82} Bott, Raoult and Tu, Loring W. 1982 , {\sl Differential Forms in Algebraic Topology}, Springer-Verlag, New York Heidelberg Berlin.Academic Press, 
\bibitem{Chevalley46} Chevalley, Claude. 1946. {\sl Theory of Lie Groups I}, Princeton University Press; Asian text edition, 1965, Overseas Publications LTD. (Kaigai Shuppan Boeki K.K.), Tokyo. 

\bibitem{Porteous95} Porteous, Ian R. 1995. {\sl Clifford Algebra and the Classical Groups}, Cambridge University Press.
\bibitem{BBJ81} Becher, Peter , Boehm, Manfred and Joos, Hans. 1981. {\sl Eichtheorien der starken und electromagnetischen Wechselwirkung}, B.G. Teubner, Stuttgart. 
\bibitem{BR06} Becchi Carlo M. and Ridolfi, Giovanni 2006, {\sl An Introduction to Relativitic Processes and the Standard model of Electroweak Interactions}, Springer-Verlag Italia.

\bibitem{FS91} Faddeev, L.D. and Slavnov, A.A. 1990. {\sl Gauge Fields -Introduction to Quantum Theory-}, Translated from the Russian Edition by Pontecorvo, G.B., Addison Wesley Publishing Company, Redwood City, California.
\bibitem{HT92} Henneaux, Marc and Teitelboim, Claudio 1992. {\sl Quantization of Gauge Systems}, Princeton University Press.
\bibitem{vanBaal92} van Baal, Pierre. 1992. {`` More (Thought on) Gribov Copies''}, Nucl. Phys, {\bf B369} 259
\bibitem{vBC92} van Baal, Pierre and Cutkosky, R.E. 1993. {``Non-Perturbative Analysis, Gribov Horizons and the boundary of the Fundamental Domain''}, Int. J. Mod. Phys. A  (Proc. Suppl.) 3A 323; 21st Conference on Differential Geometric Methods in Theoretical Physics (XXI DGM 1992)
\bibitem{vBHD92} van Baal, Pierre and Hari Dass, N.D. 1992. {``The theta dependence beyond steepest descent''}, Nucl. Phys. {\bf B385} 185-226.
\bibitem{vBvdH94} van Baal, Pierre and van den Heuvel, Bas 1994. {``Zooming-in on the SU(2) fundamental domain''}, Nucl. Phys. {\bf B417} 215-237.
\bibitem{tHooft76} t'Hooft, G. 1976. {``Computation of the quantum effects due to a four-dimensional pseudoparticle''}, Phys. Rev. {\bf D 14} 3432. Errata 1978 Phys. Rev. {\bf D18} 2199
\bibitem{Kuipers99} Kuipers, Jack B. 1999. {``Quaternions and Rotation Sequences''} Geometry, Integrability and Quantization. Varma Bulgaria. Ed. by Mladenov, I.M. et al. Coral Press, Sofia 2000, pp. 127-143.
\bibitem{Hughes00} Hughes, Noel H. 2000. {``Quaternion to Euler Angle Conversion for Arbitrary Rotation Sequence Using Geometric Methods''} www. eucledian.com
\bibitem{RRPJ05} Ramella-Roman, Jessica C., Prahl, Scott A. and Jacques, Steve I. {``Three Monte Carlo programs of polarized light transport into scattering media: part I''}, Optical Society of America.
\bibitem{Arnold78} Arnold, V.I. 1978. {\sl Mathematical Methods of Classical Mechanics}, Springer-Verlag, New York, Heidelberg, Berlin.
%\bibitem{AKN88} Arnold, V.I. , Kozlov, V.V. and Neishtadt, A.I.  1988{\sl Dynamical Systems III}, 
\bibitem{GM86} Guillou, Lucien and Marin, Alexis 1986. {\sl A la Recherche de la Toplogie Perdue}  Du cot\'e de chez Rohlin, {``Quartre articles de V.A. Rohlin''} (Translated from Russian original version by Ochanine, S..)  p.3-24. Birkhaeuser, Boston, Basel, Stuttgart.
\bibitem{Rohlin52} Rohlin, B.A. 1952. {``New Results in the 4-dimensional Manifolds''}, (In Russian) Dokladi Academy Nauk CCCP {\bf 84} (2) p. 221-224.
\bibitem{Pontryagin87} Pontryagin, L.S. 1987. {\sl Pontryagin Selected Works in Four Volumes}, Classics of Soviet Mathematics, vol. 3 {``Algebraic and Differential Topology''} p.211-249.
Mathematical Aspects of Classical and Celestial Mechanics, Chapter 1, Section2. (Translated from the Russian Eddition.) Springer-Verlag Berlin Heidelberg.
\bibitem{Furui19} Furui, Sadataka 2020. {``A Closer Look at Gluons''} Chapter 6 of  ``Horizon in World Physics vol.{\bf 302}'' , Ed. by Reimer, Albert, Nova Science Pub. 
%\bibitem{DSBM04}Dos Santos, Serge and Bou Matar, Olivier, 2004 {``Symmetry of KZ (Khoklov-    
%Zabotskaya)  equation''}, unpublished.
\bibitem{FdS19} Furui, Sadataka and Dos Santos, Serge 2019. {``Theoretical Study of Memristor and Time Reversal based Nonlinear Elastic Wave Spectroscopy''}, to be published.
\bibitem{Hoermander97} Hoermander, Lars 1997. {\sl Lectures on nonlinear hyperbolic differential equations}, Springer, Berlin.
\bibitem{DNG15} De Nittis, Giuseppe and Gomi, Kiyonori 2015, {``Classification of ``Quaternionic'' Bloch-Bundles};
Topological Quantum Systems of Type AII, Commun. Math. Phys. {\bf 339}, 1-55.
\bibitem{KR90b} Koecher, M. and Remmert, R. 1990. {``Hamilton's Quaternion'', The Isomorphism Theorems of Frobenius, Hopf and Gelfand-Mazur'', Chapter 8 of {\sl Number}, 2nd English Edition}, Translated from german original version ``Zahlen'', Springer-Verlag, New York.
\bibitem{Hatcher02} Hatcher, Allen 2002. {\sl Algebraic Topology}, Cambridge University Press, Cambridge.
\bibitem{Jimbou90} Jimbou, Michio 1990. {\sl Quantum groups and Yang-Baxter equation} (In Japanese), Springer- Verlag, Tokyo
\bibitem{Drinfeld85} Drinfeld, V.G. 1985. {``Hopf algebras and the quantum Yang-Baxter equation}. Soviet Math. Doklady {\bf 32} 254-258.
\bibitem{WZ90} Wess, Julius and Zumino, Bruno 1990. {``Covariant Differential Calculus on the Quantum Hyperplane''}, Nuclear Physics {\bf B}(Proc. Suppl.) 18B, 302-312
\bibitem{PW90} Podles, P. and Woronowicz, S.L. 1990. {``Quantum Deformation of Lorentz Group''}, Commun. Math. Phys. {\bf 130}, 381-431.
\bibitem{dTB09} de T\'eramond, Guy F. and Brodsky, Stanley 2009, {``Light-Front Holography: First Approximation to QCD''}, Phys. Rev. Lett. {\bf 102}, 081601.
\bibitem{DdTB15} Dosch, Hans Guenter, de T\'eramond, Guy F. and Brodsky, Stanley J. 2015. {``Supersymmetry Across the Light and Heavy-Light Hadronic Spectrum''}, arXiv:1504.05112v2 [hep-ph].
\bibitem{BdTDE15} Brodsky, S.J., de T\'eramond, G.F., Dosch, H.G. and Erlich, J. 2015 {``Light-front holographic QCD and emerging confinement''}, Phys. Rept. {\bf 584} 1-106.
%\bibitem{LR84} Lapidus, Yu.R. and Rudenko, O.V. 1984 {\it New approximations and results of the theory of nonlinear acoustic beams}, Sov. Phys. Acoust. {\bf 30} (6).
%\bibitem{LR92} Lapidus, Yu.R. and Rudenko, O.V. 1992 {\it An exact solution of the Khokhlov-Zaboltskaya equation}, Sov. Phys. Acoust. {\bf 38} (2).

%\bibitem{VG94} Vershik, A.M. and Gershkovich, V. Ya. 1994. {\sl Dynamical Systems VII}, I. Nonholonomic Dynamical Systems. Geometry of Distributions and Variational Problems. (Translated from the Russian Edition.) Springer-Verlag Berlin Heidelberg.
\end{thebibliography}
\end{document}